\documentclass[a4paper,12pt]{article}
\usepackage{epsfig}
\usepackage{epstopdf}
\usepackage{cite}
\usepackage[sumlimits,intlimits,namelimits]{amsmath} 

\usepackage[english]{babel}
\usepackage{color}
\usepackage{cite}

\let\oldsqrt\sqrt
\def\sqrt{\mathpalette\DHLhksqrt}
\def\DHLhksqrt#1#2{%
\setbox0=\hbox{$#1\oldsqrt{#2\,}$}\dimen0=\ht0
\advance\dimen0-0.2\ht0
\setbox2=\hbox{\vrule height\ht0 depth -\dimen0}%
{\box0\lower0.4pt\box2}}

\usepackage{geometry}
\numberwithin{equation}{section}
\usepackage{amssymb,amsmath,units}
\usepackage{color}
\usepackage{graphicx}
\usepackage{hyperref}
\usepackage{float} 
\usepackage[caption=false]{subfig}
\usepackage[T1]{fontenc}

\begin{document}
\begin{titlepage}
\begin{flushright}
TUM-HEP-1259/20\\[0.2cm]
SI-HEP-2020-06\\[0.2cm]
\today
\end{flushright}

\vspace{1.2cm}
\begin{center}
{\Large\bf 
CP Violation in Three-body \boldmath $B$ \unboldmath Decays \\[2mm] 
A Model Ansatz}
\end{center}

\vspace{0.5cm}
\begin{center}
{\sc Thomas Mannel, Kevin Olschewsky} \\[0.1cm]
{\sf Theoretische Physik 1, Naturwiss. techn. Fakult\"at, \\
Universit\"at Siegen, D-57068 Siegen, Germany}\\[0.2cm]
{\sc  K. Keri Vos} \\[0.1cm]
{\sf Physik Department, Technische Universit\"at M\"unchen, \\ James-Franck-Str.~1, D-85748 Garching, Germany}
\end{center}

\vspace{0.8cm}
\begin{abstract}
\vspace{0.2cm}\noindent


The mechanism of CP violation remains one of the puzzles in particle physics. Key to understanding this phenomenon are nonleptonic $B$ decays, especially multibody decays which exhibit large CP asymmetries in various regions of phase space. A full QCD-based theoretical description of these decays is still missing, requiring the use of models to fit the data. In this paper, we suggest a model ansatz which reflects the underlying physics and the known mechanism of CP violation via the CKM matrix. In addition, since CP violation is driven by the interference between amplitudes with and without valence charm quarks, we argue that the opening of the open-charm threshold may play an important role in generating CP violation in the high invariant mass region. We present a natural extension of the isobar model to incorporate these effects. We suggest an analysis of nonleptonic three-body $B$ decay data including this extension, which would be interesting as it may give new hints to the sources of CP violation both at low and high invariant mass.

\end{abstract}

\end{titlepage}

\section{Introduction}
The violation of CP symmetry is one of the main puzzles of contemporary particle physics. 
In the Standard Model (SM) of particle physics, CP violation in the electroweak interaction is described by a complex parameter in the quark-mixing (CKM) matrix. On the other hand, the strong CP problem, namely 
the absence of (flavour diagonal) CP violation in strong interactions, remains a mystery. Flavour physics experiments over the last two decades confirm this picture, which is, however, not a theory of CP violation but rather a parametrization of the observations. 

An important source of information on CP violation are $B$ decays, for which large amounts of data are available from BaBar, Belle(II) and LHCb. For the charmless nonleptonic two-body decays the method of QCD factorization (QCDF) \cite{Beneke:2000ry, Beneke:1999br, Beneke:2001ev, Beneke:2003zv} has been established, which is set up as a double expansion in $\alpha_s$ and $\Lambda_{\rm QCD}/m_b$ and allows the computation of branching ratios and CP asymmetries. 

A large part of the nonleptonic $B$ decays are however three- and more-body decays. These have the added benefit that the strong phases and therefore also the CP asymmetries differ across the phase space.  Recently, the QCDF framework has been applied to three-body $B$ decays \cite{Krankl:2015fha, Klein:2017xti}. First attempts to calculate decay rates and CP asymmetries within this framework show that the leading order in QCDF can reproduce the gross features for the Dalitz distributions for differential rates, but do not seem to describe the corresponding CP asymmetries. This may indicate the importance of subleading power corrections, for which currently no systematic description exists. 

Until a sound theoretical approach exists, the description of multibody $B$ decays relies to a large extent on modelling (see e.g. \cite{Cheng:2007si,Bediaga:2009tr, Nogueira:2015tsa, Cheng:2016shb,Bhattacharya:2013boa,Boito:2017jav,  Bediaga:2017axw, Bediaga:2018wml,Bediaga:2020ztp}). Therefore, we revert the usual arguments. Instead of using data to extract the parameters of CP violation, we assume that CP violation is described by the SM. In that way, the measured CP asymmetries can be used to determine the hadronic matrix elements and improve our understanding of QCD. 

In this paper, we propose a new parametrization for nonleptonic three-body $B$ decay amplitudes that reflects the SM mechanism of CP violation by dividing the amplitude into ${\cal A}_u$ and ${\cal A}_c$, which contain valence $u$ and $c$ quarks, respectively. CP violation is driven by the interference between these amplitudes. At high invariant mass, dominant contributions to ${\cal A}_c$ could be charm resonances, but also charmonium-like exotic states. These contributions may be described by the standard $2+1$ description, where the decay is described as a pseudo two-particle decay. Besides such effects, threshold singularities, for example generated by the opening of the $D\bar{D}$ threshold may constitute a large part of ${\cal A}_c$ and thus play an important role in the description of CP violation, as we already pointed out in \cite{Klein:2017xti}. Such threshold effects in three-body $B$ decays were recently studied in Ref.~\cite{Bediaga:2017axw, Bediaga:2018wml, Bediaga:2020ztp} using a mesonic model, which generates a sharp strong phase change when crossing the $D\bar{D}$ threshold.

The key point of the current paper is to emphasize the importance of charmed resonances and threshold effects. As such effects are challenging to describe in QCD, we propose a simple extension of the standard isobar model to parametrize charm threshold effects. Using a simple example, we show the CP violation such threshold effects could generate. Whether these effects play indeed an important role can only be determined using a full analysis of the three-body $B$ decay data using our new model ansatz. This in turn would then generate new insights which would help to further improve the parametrization and allow us to gain insights into charm effects in nonleptonic $B$ decays.

\section{Modelling the three-body amplitudes} 
We focus on $B^+(p_B) \to \pi^+(p_1) \pi^-(p_2) \pi^+(p_3)$, but our approach can easily be extended to $B\to hhh$ with $h = \pi, K$ decays. 
In the SM, the decay is described by the weak effective Hamiltonian for the 
$b \to d$ flavour transition \cite{Buchalla:1995vs}, which can be split into two RG invariant contributions with different CKM elements
\begin{equation}  \label{Heff1} 
    H_{\rm eff} = \frac{G_F}{\sqrt{2}}  \left[ V_{ub} V_{ud}^*  O_u   
    +  V_{cb} V_{cd}^* O_c   \right],
\end{equation} 
where $O_{u,c}$ contain current-current and 
penguin operators. Using the convention that the CP-violating weak phase of the CKM matrix enters via $V_{ub} = | V_{ub} | e^{-i\gamma}$, the decay amplitude can be written as
\begin{equation}
    {\cal A}_\pm  (s_{12}, s_{23})  \equiv 
    \langle B^\pm | H_{\rm eff} |  \pi \pi \pi \rangle = {\cal A}_u (s_{12}, s_{23})\, e^{\mp \mathrm{i}\gamma} + {\cal A}_c (s_{12}, s_{23})  
\end{equation} 
with $ {\cal A}_q (s_{12}, s_{23})  = G_F/\sqrt{2}  | V_{qb} V_{qd}^* |  \langle B |  O_q  |  \pi \pi \pi \rangle$. 
The amplitudes ${\cal A}_\pm $, ${\cal A}_u $ are complex valued functions of the kinematic variables $s_{12}$ and $s_{23}$, where $s_{ij}\equiv m_{ij}^2 = (p_i + p_j)^2$, and thus contain CP-conserving strong phases. Direct CP violation is induced by the interference of 
the matrix elements of the two operators $O_u$ and $O_c$. 

The matrix elements ${\cal A}_q$ cannot be calculated reliably, therefore their values have to be extracted from data by means of an amplitude analysis. In general, ${\cal A}_+ $ and ${\cal A}_- $ are fitted separately using a model ansatz. Frequently used are isobar (inspired) models, which parametrize the three-body amplitudes as pseudo two-particle decays, where one of the two particles is a resonance that subsequently decays into two stable particles. Schematically;
\begin{equation} \label{isomod}
    {\cal A}_\pm(s_{12},s_{23}) =\sum_k \frac{c_k^\pm P_k^{(\ell)} (s_{12},s_{23}) }{s_{12}-m_k^2 + i m_k \Gamma_k} 
    + \sum_l \frac{c_l^\pm P_l^{(\ell)} (s_{12},s_{23}) }{s_{23}-m_l^2 + i m_l \Gamma_l}.
\end{equation}
Here $P^{(\ell)}$ are the Legendre polynomials describing the spin of the decaying resonances with masses $m_{k/l}$ and widths $\Gamma_{k/l}$. In the simplest isobar model the propagators of the decaying resonances are described by Breit-Wigner shapes. The model contains the complex parameters $c_k^\pm$ which are assumed to be constant, such that all the kinematic dependence arises from the Breit-Wigner form. Thus the strong phases extracted from the Dalitz distribution for the differential rates depend strongly on the model assumption used to fit the data.

A similar approach is used in $D$ decays, where CP violation can safely be ignored in the amplitude analysis. However, in $B$ decays there is additional information through the CP distributions. As the weak CKM phase $\gamma$ is constant throughout the Dalitz plane, the CP distribution gives a direct measure of the strong phase differences between the amplitudes ${\cal A}_u $ and ${\cal A}_c$. This difference is driven by the mass of the charm quark, if the charm and up-quark mass would be equal, the CP asymmetry would vanish, since then ${\cal A}_u $ and ${\cal A}_c $ become identical. 

In the isobar model, the kinematical dependence of the phases in ${\cal A}_u$ and ${\cal A}_c$ arise from the Breit-Wigner shapes and are related to the asymptotic final-state interactions of the decay 
products of the resonance. These asymptotic interactions are not related to CP violation, which is 
evident since both interfering amplitudes will have the same Breit-Wigner phase. Therefore, the fact that the thresholds for charm states are fundamentally different from light-quark states will manifest itself in the CP asymmetry close the charm thresholds. Such effects would contribute to ${\cal A}_c$, while they would be absent (or strongly suppressed) in ${\cal A}_u$. 
Interestingly, the distribution of CP violation for $B\to\pi\pi\pi$ obtained by the LHCb Collaboration \cite{Aaij:2019jaq, Aaij:2019hzr,Aaij:2014iva} shows pronounced structures in the center 
of the Dalitz plot close to the invariant mass of $c \bar{c}$ states.

Therefore, the parameterization of the amplitudes in ${\cal A}_u $ and ${\cal A}_c$ seems a natural choice in which the known mechanisms of CP violation in the SM are incorporated. We therefore propose to perform the analysis using not the amplitudes ${\cal A}_+ $ and ${\cal A}_- $ but rather ${\cal A}_u $ and ${\cal A}_c$. Specifically this implies
\begin{align}\label{eq:isobar}
        {\cal A}_\pm  (s_{12}, s_{23}) =& 
        \sum_k \frac{(a_k^{(u)} e^{\mp i\gamma} + a_k^{(c)}) P_k^{(\ell)}  (s_{12}, s_{23}) }{s_{12}-m_k^2 + i m_k \Gamma_k} 
        + \sum_l \frac{(b_l^{(u)} e^{\mp i\gamma} + b_l^{(c)}) P_l^{(\ell)}  (s_{12}, s_{23}) }{s_{23}-m_l^2 + i m_l \Gamma_l} \ , 
\end{align}
with the same set of complex fit parameters $a^{(u)}$ and $a^{(c)}$ for both amplitudes. 

In addition, the benefit of a set up using $\mathcal{A}_u$ and $\mathcal{A}_c$ is, that charm effects which are expected to play a dominant role above the open-charm threshold can be included systematically. We discuss the inclusion of these threshold effects in the next section.

We note that the isobar-model parametrization has a few interesting features. First of all,
in the case of very narrow resonances (meaning that their width is small compared to the mass spacing 
between the different resonances) the relation in Eq.~\eqref{eq:isobar} implies
\begin{eqnarray}
\frac{d^2 \Gamma_{\pm}}{ds_{12} \, ds_{23}} &=& \sum_k \delta(s_{12} - m_k^2) P_k^{(\ell)} (m_k^2,s_{23}) {\rm Br} (B^\pm \to \pi R_k)  
\end{eqnarray}  
where 
\begin{equation}
{\rm Br} (B^\pm \to \pi R_k) = \frac{1}{\Gamma_{\rm tot}} | a_k^{(u)} e^{\mp i\gamma} + a_k^{(c)} |^2 \ ,
\end{equation}
is the branching ratio of the two-body decay $B\to \pi R_k$. 
In this approximation, the rate asymmetry is thus completely determined by those of the two particle decays. Therefore, if only low lying resonances 
with masses well below the charm threshold are take into account, the narrow width example clearly shows that the complex 
structure in the $CP$ asymmetries at high-invariant masses can hardly be accounted for. 

A second remark needs to be made concerning the current fits of the data. In the most recent analysis of the LHCb Collaboration \cite{Aaij:2019hzr, Aaij:2019jaq} the residues of the resonances are parametrized as
\begin{equation}\label{eq:para}
 c_k^\pm = a_k^{(u)} e^{\mp i\gamma} + a_k^{(c)} = x_k \pm \delta x_k + i (y_k \pm \delta y_k) \ ,
\end{equation}  
with real parameters $x_k, y_k, \delta x_k, \delta y_k$ and where $a^{(q)} = |a^{(q)}| e^{i \phi^{(q)}}$. Equation~\eqref{eq:para} shows that the two parameterizations are related and contain the same information. We emphasize, however, that when using $x,y,\delta x, \delta y$ the assignment to the matrix elements appearing in the effective 
Hamiltonian is not evident any more. In practice, the fit values are extracted with respect to the $\rho$ resonance \cite{Aaij:2019hzr, Aaij:2019jaq}. This means that $y_\rho$ and $\delta y_\rho$ are fixed to $0$ corresponding to a phase choice, because the overall phase of amplitude ${\cal A}_\pm$ is not observable. For our parametrization this implies that $\phi_\rho^{(u)}=90^\circ$. In addition, for the isobar fit $x_\rho$ is fixed to $1$, so the values of $x,y, \delta x, \delta y$ for other resonances are in units of the value for the $\rho$ resonance. The relative strength of two matrix elements ${\cal A}_u$ and ${\cal A}_c$ is then given by
\begin{equation}\label{eq:ratio}
    \frac{|a_\rho^{(u)}|}{|a_\rho^{(c)}|} \simeq \delta x_\rho \simeq {\cal O}(10^{-3}) \ ,
\end{equation}
where the numerical factor can be obtained by using the fit result for $\delta x_\rho$ from \cite{Aaij:2019hzr}. The smallness of the ratio seems to indicate that the amplitude for the $\rho$ resonance is driven by ${\cal A}_c$. This seems counter-intuitive as the $\rho$ could be immediately formed from the valence $u$ quarks in ${\cal A}_u$. Therefore, we would expect the ratio in Eq.~\eqref{eq:ratio} to be much bigger. The small ratio may very well be an artefact of the analysis as charmed resonances are not included. In fact, rewriting the fit parameters $x,y,\delta x, \delta y$ from \cite{Aaij:2019hzr} for the $\sigma$ $S$-wave and $f_2$ $D$-wave resonance leads also to a small ratio $a^{(u)}/a^{(c)}$.  This could be an indication that charmed resonances and threshold effects play a role. Finally, we note that the integrated CP asymmetry in the high-$s$ region seems well described by the LHCb amplitude fits \cite{Aaij:2019hzr}, however, it would be interesting to see if the amplitude analysis can reproduce the full CP landscape. 

In the next section, we suggest a simple parametrization to include also charm threshold effects in the isobar fit. We expect that the proper inclusion of charm threshold effects will change the ratio in Eq.~\eqref{eq:ratio} to more plausible values.

\section{Parametrization of threshold effects}
To parametrize the impact of the open-charm threshold effects, we concentrate on the current-current part of the operator $O_c$, which we write as
\begin{equation} \label{sum}
    \langle B^+ |  O_c  |  \pi \pi \pi \rangle = \sum_n 
    \langle B^+ |  O_c  | n \rangle \, \langle n | \pi \pi \pi \rangle,
\end{equation}
where $| n \rangle$ is an intermediate state. The sum over these intermediate states can be split into the states with and without valance charm quarks. We will assume that the difference between the matrix elements of the operators $O_c$ and $O_u$ is mainly driven by the states involving valence charm quarks and hence focus on them. 

We naively factorize the matrix element  $ \langle B^+ |  O_c  | n \rangle$, by fierzing the four-quark operator as 
\begin{eqnarray}
\langle O_c \rangle = \frac{1}{N_c} 
    \langle \bar{b} \gamma_\mu (1-\gamma_5) d \rangle \langle \bar{c} \gamma_\mu (1-\gamma_5) c \rangle   +\langle  
    \bar{b} \gamma_\mu T^a (1-\gamma_5) d \rangle \langle\bar{c} \gamma_\mu T^a (1-\gamma_5) c \rangle \ .
    \label{Fierz}
\end{eqnarray}  

The invariant mass of the intermediate states $| n \rangle$ must be close to the $B$ meson mass, and therefore the lowest possible (hadronic) state with valence charm quarks is a $c \bar{c}$ resonance $R_{c\bar{c}}$ and a pion. At leading order in the Fock state, the $c \bar{c}$ resonance is in a color singlet state, thus only the first term in Eq.~\eqref{Fierz}, which is $1/N_c$ suppressed, contributes. Furthermore, the $c \bar{c}$ resonance $R_{c\bar{c}}$ is in a $V-A$ isopin-zero state. The (axial)-vector current generates a tower of $J/\psi$ ($1^{--}$) (and orbitally excited charmonium) resonances which decay into two pions, but only with small branching ratios. 

\begin{figure}[t]
	\centering
	\subfloat[]   {\includegraphics[width=0.3\textwidth]{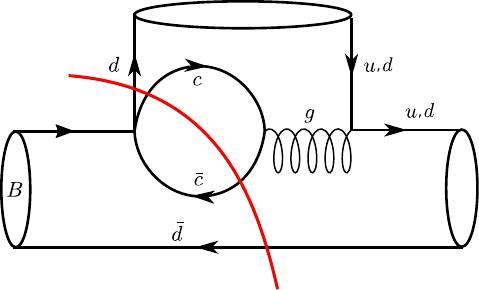}}\hspace{1cm}
	\subfloat[]{   \includegraphics[width=0.3\textwidth]{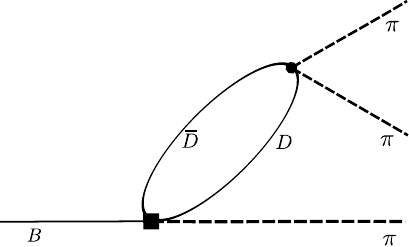}}
	\caption{(a) Sketch of color-octet contribution in QCD factorization, where the
blobs correspond a single or two-pion states (b) Example of a mesonic diagram corresponding to (a).}
	\label{fig_QCD_charm}
\end{figure}

The color-allowed second term in Eq.~\eqref{Fierz} produces the two charm quarks in a color-octet state. Its leading contribution arises through intermediate states containing a $D^{(*)} \bar{D}^{(*)}$ pair. A similar picture emerges in QCD factorization of two body decays, where the charm loop drives the CP asymmetry. This is sketched in Fig.~\ref{fig_QCD_charm}, where the cut shows that the $D \bar{D}$ states can appear as intermediate states. In two-body decays, the loop momentum $q$ has $q^2=m_B^2$ and therefore this contribution is subleading in $1/m_b$. However, in three-body decays such $D \bar{D}$ intermediate state can rescatter into two pions which leads to a threshold behaviour once the invariant mass of the two pions crosses the value $2 m_D$. In general, the amplitude 
${\cal A}_c $ will thus contain - aside from Breit-Wigner like contributions as modelled in the isobar model - also threshold-like singularities
\begin{equation}\label{eq:Ocsplit}
   {\cal A}_c =  
    \langle B^+ | O_c | R_{c \bar{c}} \pi \rangle \, 
    \langle R_{c \bar{c}} \pi |  \pi \pi \pi \rangle \\
    + \langle B^+ | O_c | D \bar{D} X \rangle \, 
    \langle  D \bar{D} X |   \pi \pi \pi \rangle,
\end{equation}
where for simplicity we only consider the $D\bar{D}$ threshold. We denote the intermediate $D\bar{D}X$-state by $R$, where $R$ contains the $c\bar{c}$ pair in a color-octet configuration. The second term in Eq.~\eqref{eq:Ocsplit} is challenging to calculate, even if one constructs a mesonic model based on heavy-meson chiral perturbation theory. The same applies to a dispersive treatment, where additional subtraction constants need to be introduced, limiting the predictive power of the approach. Therefore, we propose a simple model ansatz for these threshold contributions.

\begin{figure}[t]
	\centering
	\subfloat[]{\includegraphics[width=0.3\textwidth]{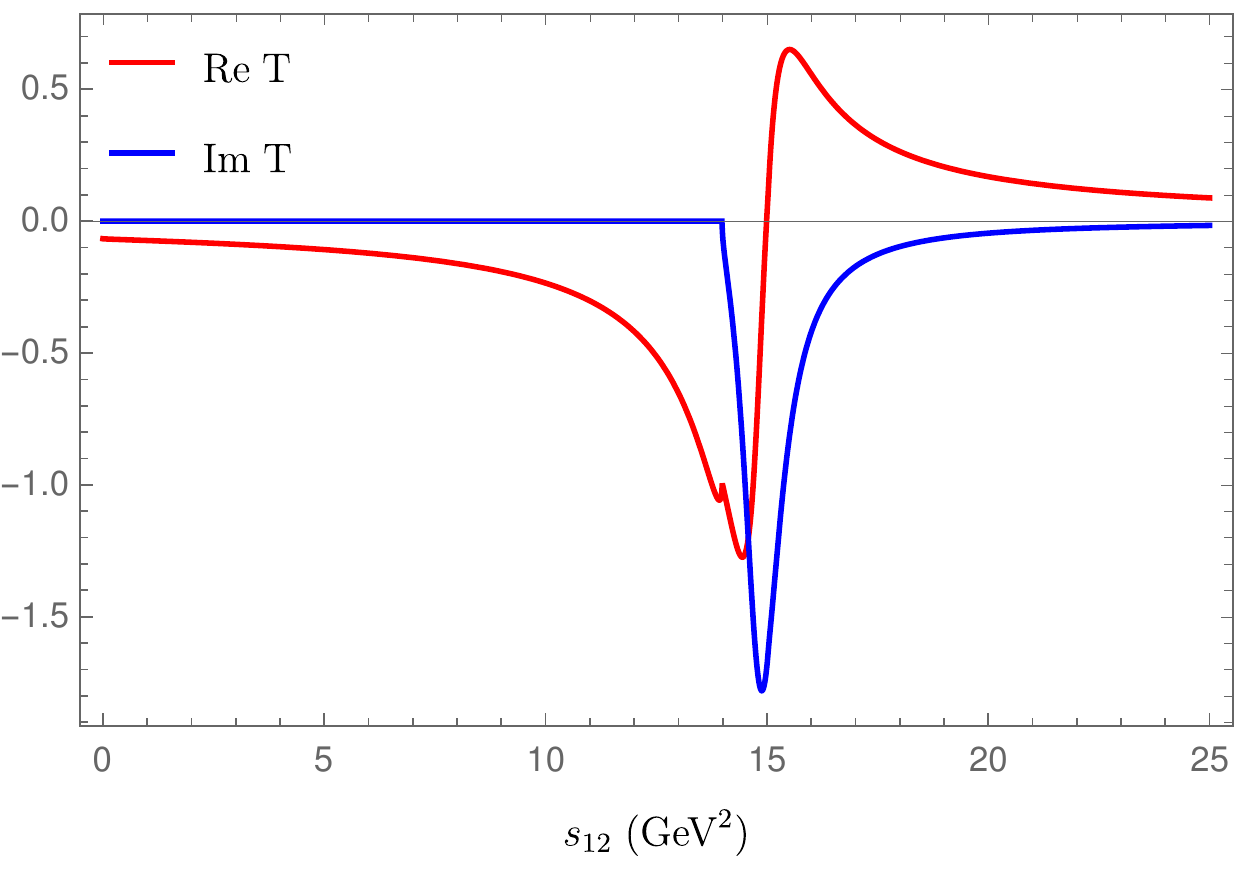}}\hspace{1cm}
	\subfloat[]{\includegraphics[width=0.29\textwidth]{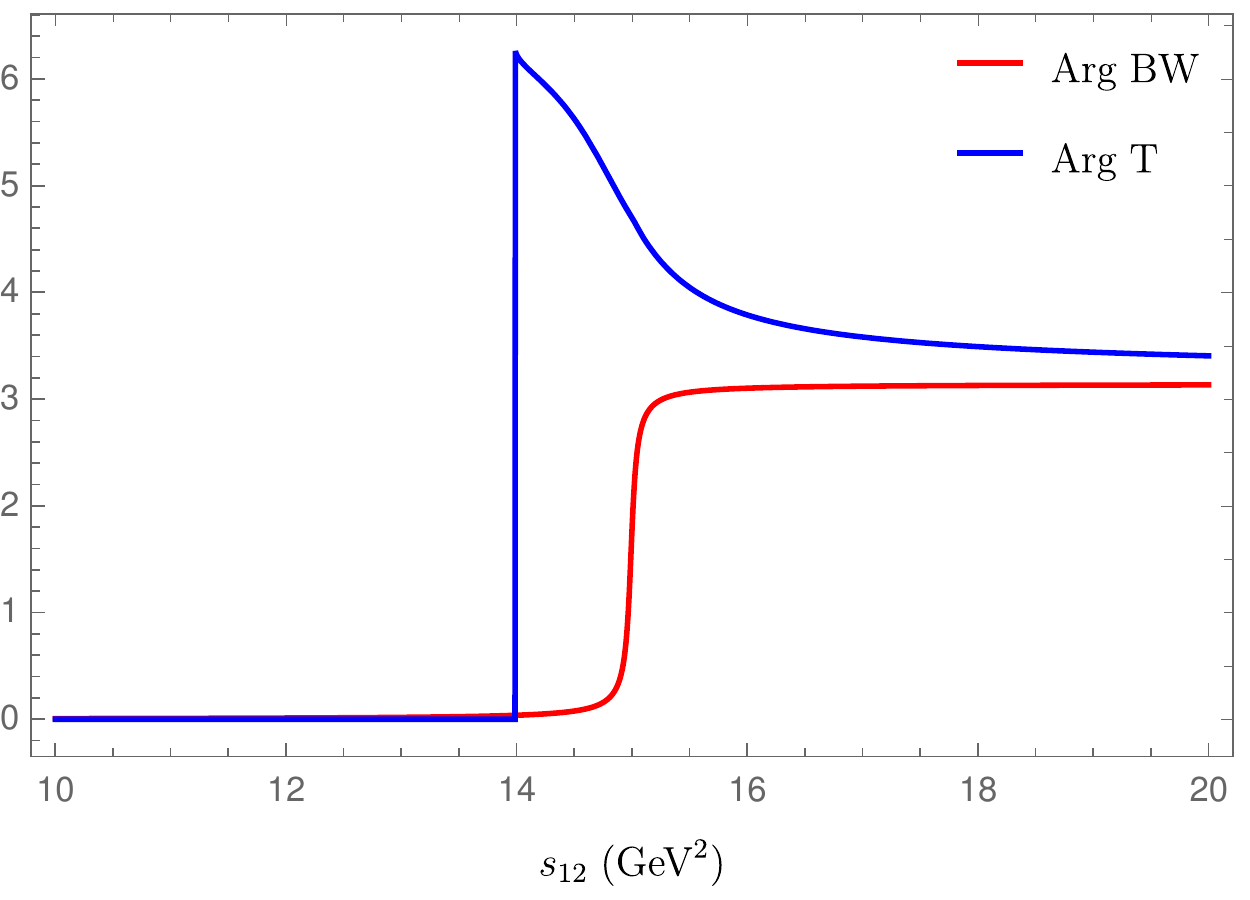}}
	\caption{(a) Real and imaginary parts of the modified propagator $T_R$ below and above the open-charm threshold. (b) Comparison of the argument of $T_R$ and a standard Breit-Wigner shape.}
	\label{fig_beh_T}
\end{figure}

The intermediate state $R$ can be described with a modified propagator corresponding to a two-point function of a quantum field
\begin{equation}
    T_R(s_{12}) = \frac{1}{s_{12} - (m_R^{{\rm bare}})^2 + \Sigma(s_{12})} = \frac{1}{s_{12} - m_R^2 +  [\Sigma_R(s_{12})- \text{Re}\;\Sigma_R(m_R^2)]} \ ,
\end{equation}
where $\Sigma_R(s_{12})$ is the self energy of the state $R$. To ensure that the pole is located at the physical mass $m_R$, the bare mass is renormalized by Re $\Sigma_R(m_R^2)$\footnote{We do not explicitly consider the field renormalization, as these terms would only alter the residue of the propagator which is a model parameter.}. Here we focus on the contributions from the open-charm loops to the self-energy, which entail the standard bubble summation diagrams. The underlying assumption of this parametrization is the fact, that the self energy of the resonance decaying into two pions may be represented by this particular threshold contribution. These loops generate a dynamical width for the resonance above the open-charm threshold, which we parametrize by the threshold function
\begin{align}
\label{eq_2pthres}
  \Sigma_R(s_{12}) =g_R m_R  \sqrt{s_\text{thres}-s_{12}}\arctan\left(\frac{1}{\sqrt{\frac{s_\text{thres}}{s_{12}} - 1 + i\epsilon}}\right) ,
\end{align}
where $s_\text{thres}$ is the open-charm threshold and $g_R$ is the coupling of the intermediate state $R$ to an open-charm system. 
Figure ~\ref{fig_beh_T} depicts the real and imaginary part of $T_R$ below and above the open-charm threshold, and a comparison between the argument of $T_R$ and a standard Breit-Wigner shape. This shows that the strong phase introduced from the threshold is fundamentally different from that in a standard Breit-Wigner parametrization. This sharp sign change at the $D\bar{D}$ threshold was also found in \cite{Bediaga:2020ztp, Bediaga:2017axw}. As a consequence, our parametrization generates much larger structures in the CP distribution. We note that in principle the coupling $g_R$ can be fixed once the width of $R$ can be determined from experiment, however, also effects of composite states could be described by Eq.~\eqref{eq_2pthres}.

Finally, the threshold effects described by $T_R$ are accounted for via
\begin{align}\label{eq:Ac}
    \mathcal{A}_c(s_{12}, s_{23}) =  \sum_R a_R e^{i\phi_R} \left( P^{(\ell)}_R(s_{12}, s_{23}) T_R(s_{12})  + (s_{12} \leftrightarrow s_{23})\right) + \ldots, 
\end{align}
where $a_R$ and $\phi_R$ are constant normalization constants and phases, and the dots indicate low-lying resonances that can be parametrized in the usual way. In principle, $m_R, g_R, a_R$ and $\phi_R$ are fit parameters that should be determined from the amplitude analysis. 

On the other hand, known exotic charmonium resonances can also be explicitly included. For example $X(3872)$ or $Z_c(3900)$, which decay predominantly into a pair of open-charm states (like $D\bar{D}$, $D\bar{D}^\ast$, $\ldots$) and lie very close to their respective thresholds. Finally, we emphasize that these $D\bar{D}$ threshold effects may be even more pronounced in $B\to KKK$ or $B\to K \pi \pi$ decays. The above discussion can be similarly applied to these decays and our model ansatz for ${\cal{A}}_c$ in Eq.~\eqref{eq:Ac} could directly be applied to these analyses as well.

\begin{figure}[t]
	\centering
	\subfloat[]{\includegraphics[width=0.5\textwidth]{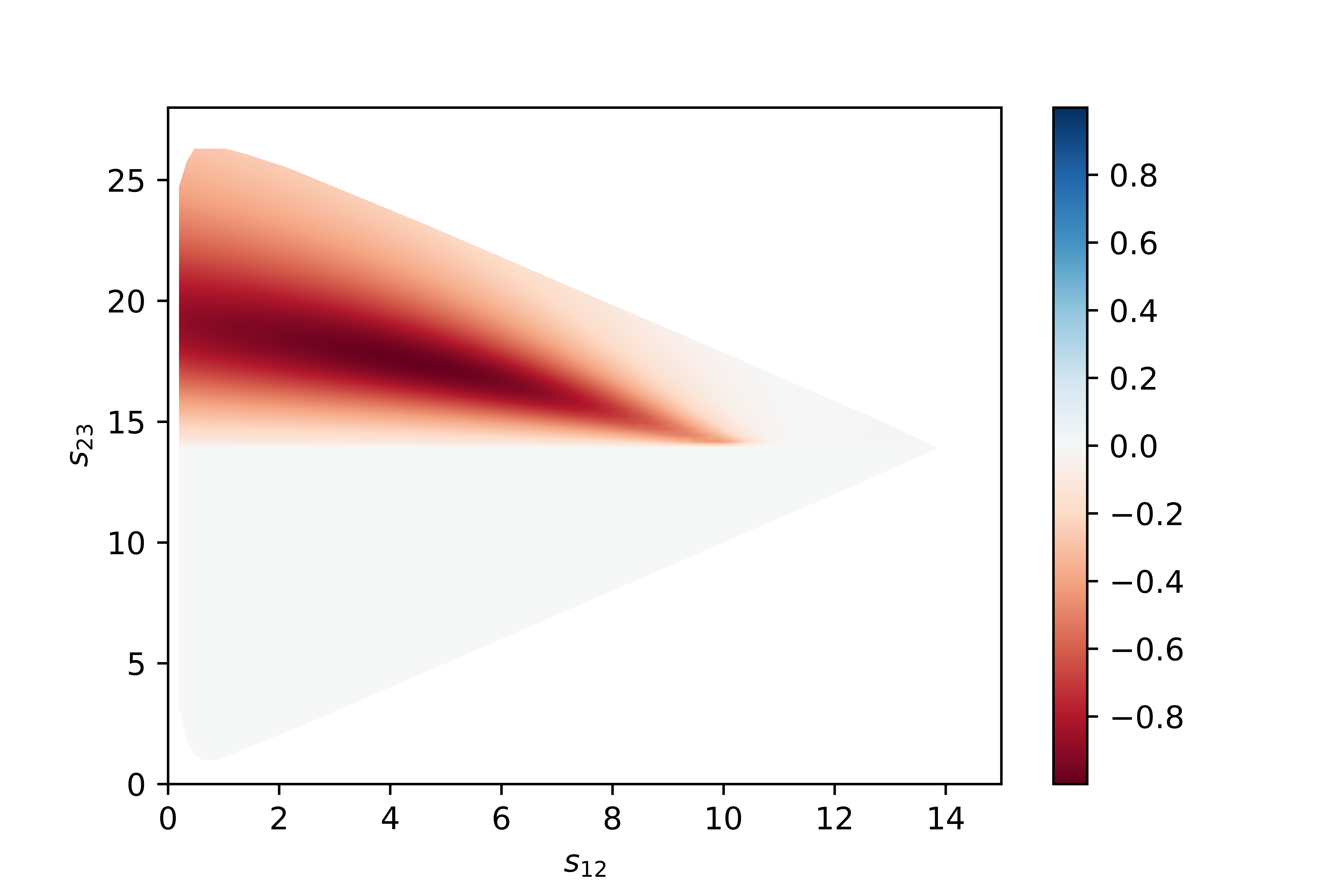}}
	\subfloat[]{\includegraphics[width=0.5\textwidth]{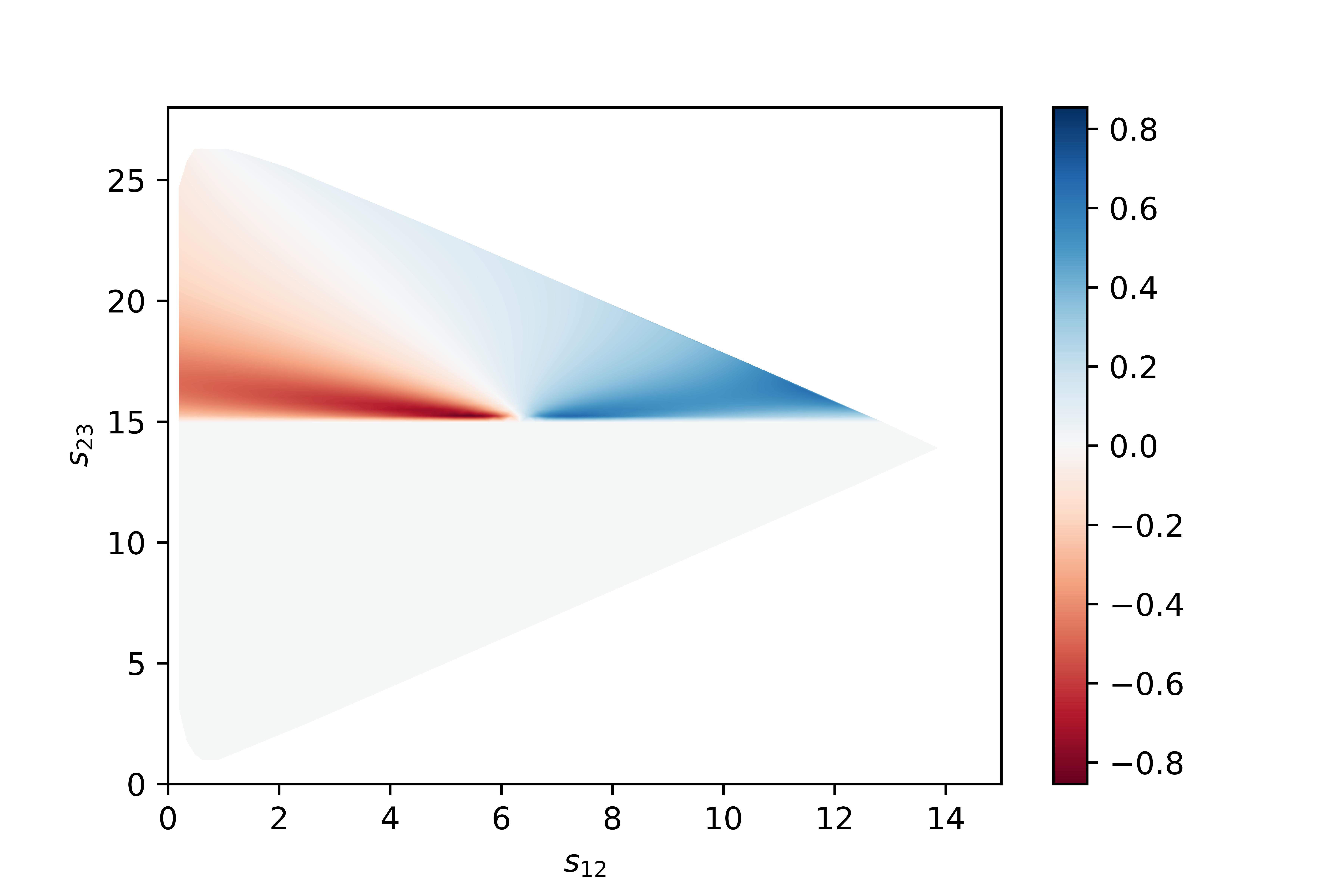}}
	\caption{CP distributions from the resummed propagator $T_R$  (a) for the $S$-wave resonance $\chi_{c0}(3860)$ (b) the $P$-wave resonance $X(3872)$. }
	\label{fig_dalitz_thres}
\end{figure} 

\begin{figure}[t]
	\centering
	\subfloat[]{\includegraphics[width=0.5\textwidth]{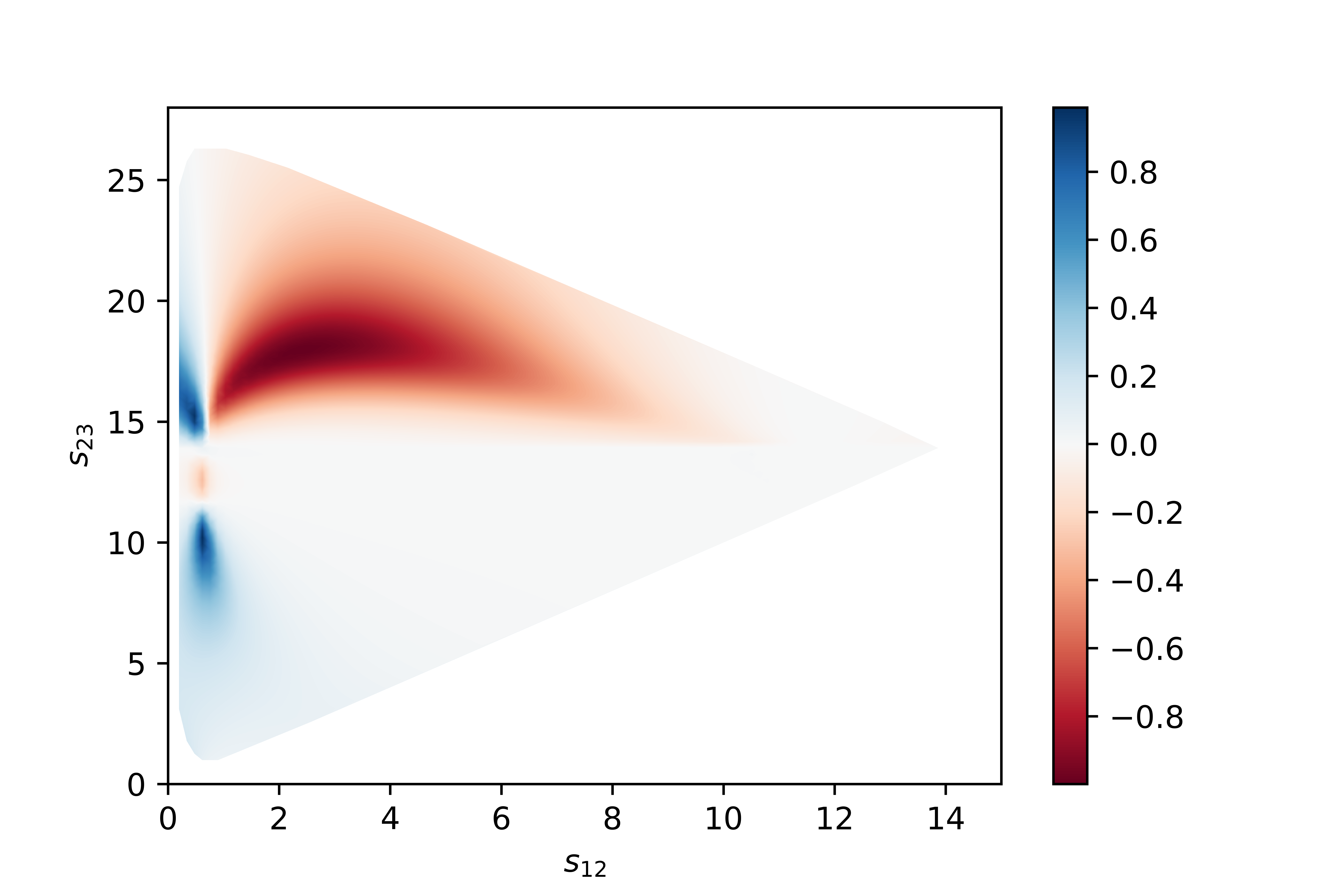}}
	\subfloat[]{\includegraphics[width=0.5\textwidth]{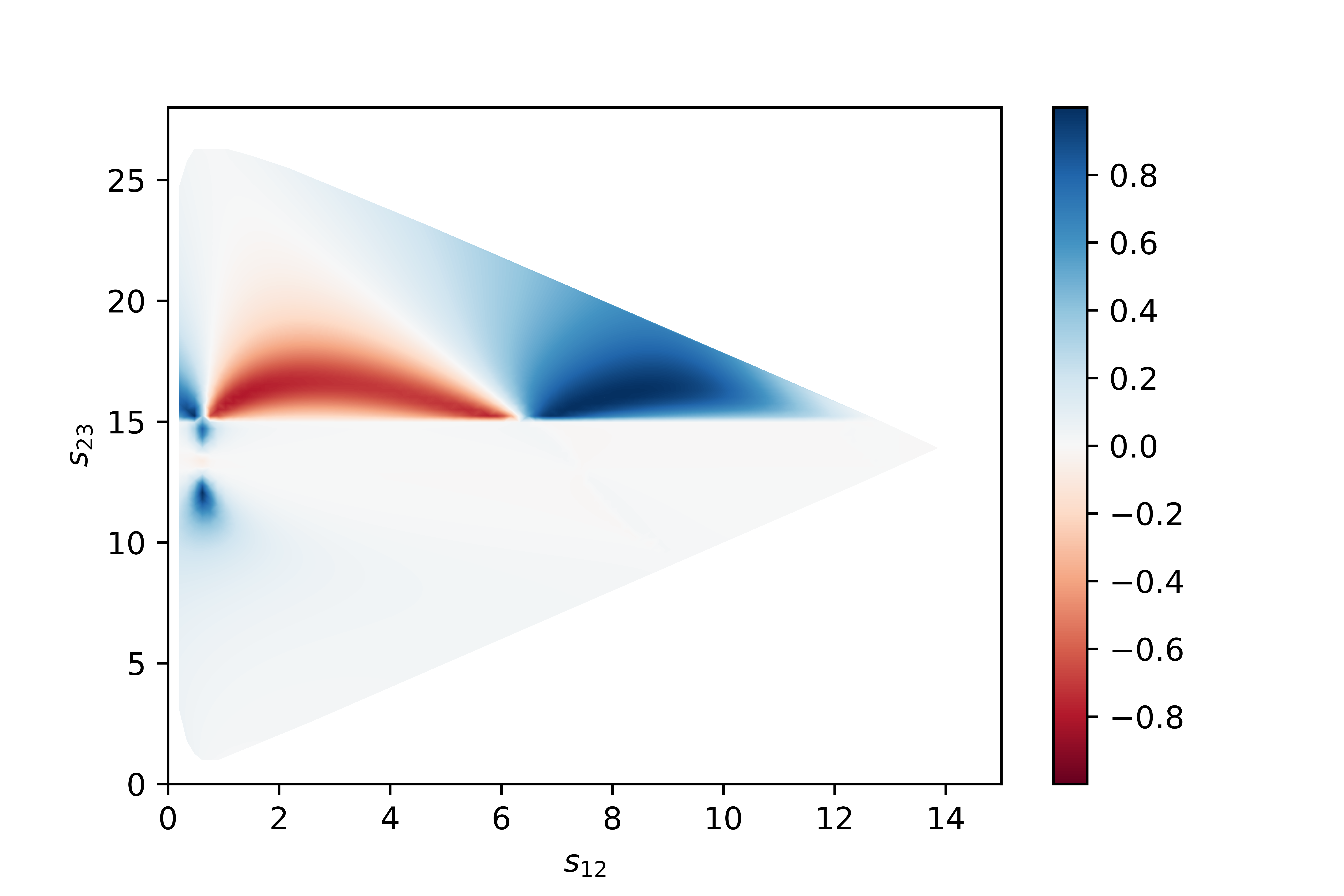}}
	\caption{CP distributions from interference of a $\rho$ resonances and the resummed propagator $T_R$ (a) for the $S$-wave resonance $\chi_{c0}(3860)$ (b) the $P$-wave resonance $X(3872)$. }
	\label{fig_dalitz_combi}
\end{figure}

\begin{figure}[t]
	\centering
	\subfloat[]{\includegraphics[width=0.5\textwidth]{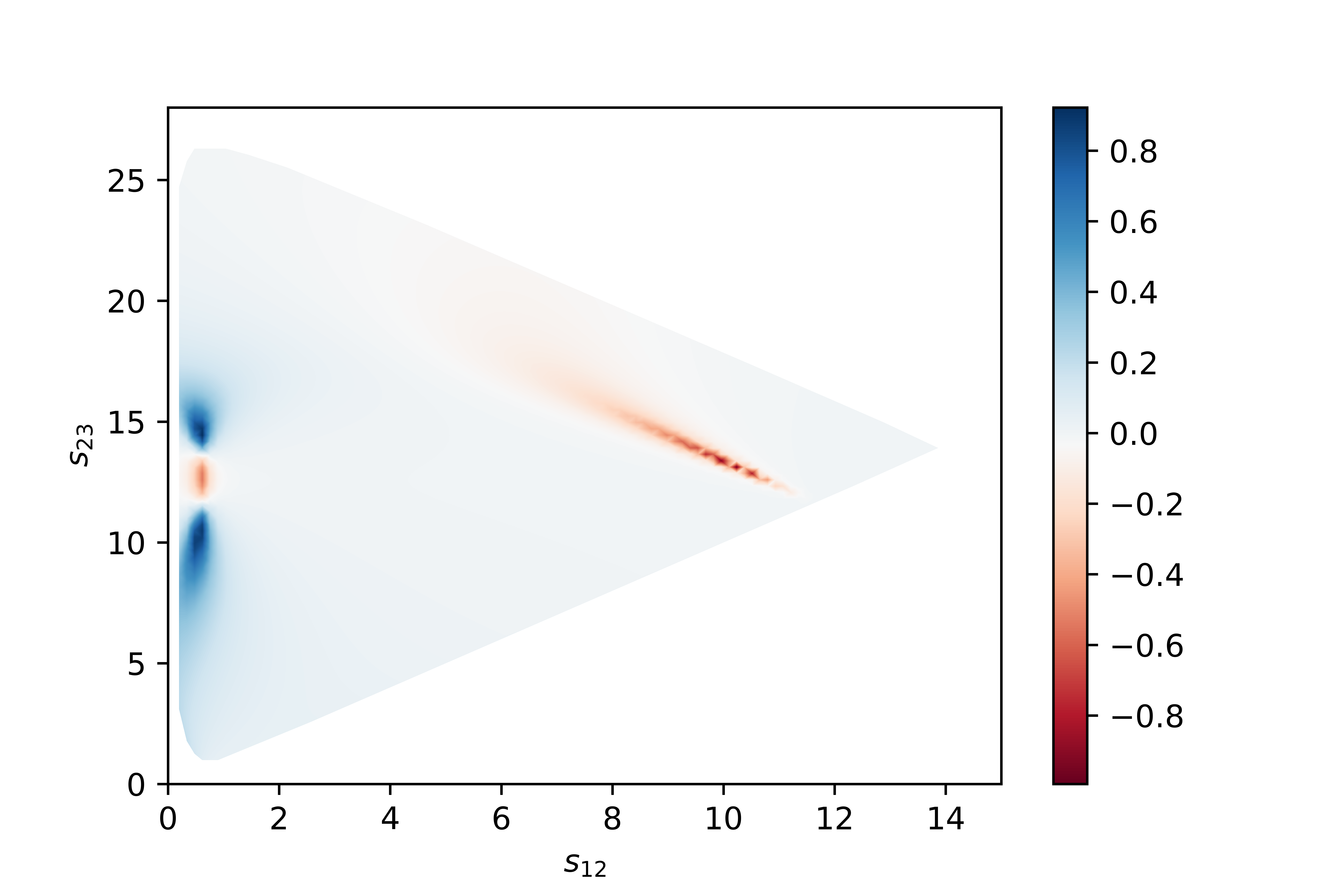}}	\subfloat[]{\includegraphics[width=0.5\textwidth]{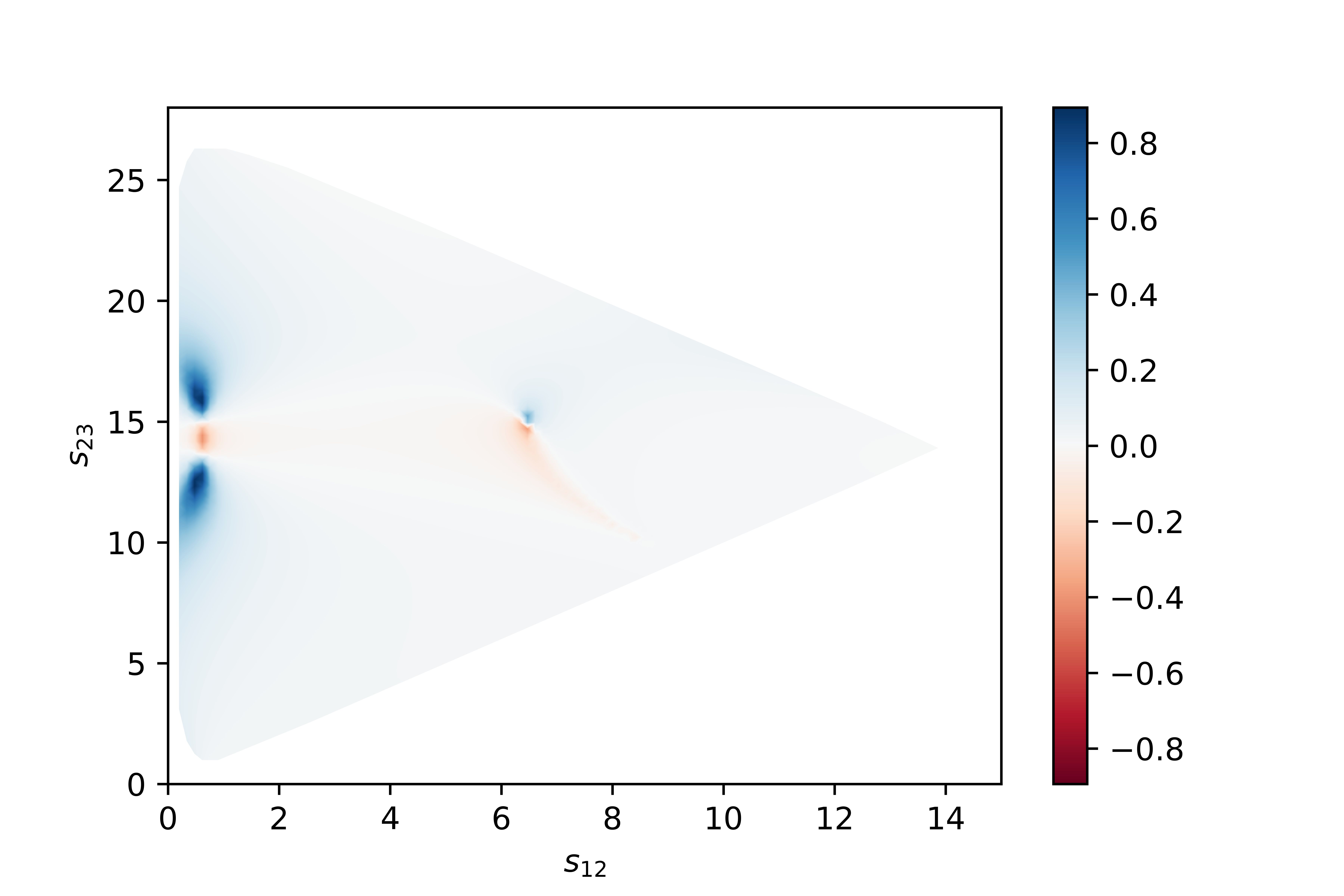}}
	\caption{CP distributions from interference of a $\rho$ resonances and Breit-Wigner parametrizations (a) for the $S$-wave resonance $\chi_{c0}(3860)$ (b) the $P$-wave resonance $X(3872)$. }
	\label{fig_dalitz_BW}
\end{figure}

\section{Discussion and Conclusion}
Our simple model ansatz for the threshold effects should be tested in a full analysis of the $B\to hhh$ data to see if the CP asymmetries in the high-energy region can be described. As discussed, it would specifically interesting to see if the fit results for the lower-lying resonances then change to be more in line with the expectation that they are dominated by ${\cal{A}}_u$. In addition, it would interesting to see these effects taken into account for the $B \to KKK$ amplitude analysis. 

A full analysis of the experimental data lies beyond the scope of this paper, as it also requires the inclusion of the lower-lying resonances and specific $S$ and $D$ wave parametrizations as done by the LHCb Collaboration (see \cite{Aaij:2019jaq, Aaij:2019hzr} for a recent elaborate study). Nevertheless, to illustrate that the sharp change in strong phase due to the opening of the $D\bar{D}$ threshold can generate interesting CP patterns, we will discuss four simple examples. 

First, we show in Fig.~\ref{fig_dalitz_thres} the effect of $T_R$ in the high-s region by assuming that ${\cal A}_u$ is constant and by considering the effect of one resonance in ${\cal A}_c$. To emphasize the difference between $S$ and $P$-wave contributions, we consider as an example the scalar $\chi_{c0}(3860)$ and the vector $X(3872)$ described by the resummed propagator $T_R$ where the relevant thresholds are $D\bar{D}$ and $D\bar{D}^\ast$, respectively. We emphasize that these states are just examples to illustrate the effect of $T_R$. In fact, our parametrization is not limited to these known resonances and can be applied more general by letting the $m_R$ and $g_R$ free in the analysis. Here, we fixed $m_R$ to the value of the respective resonance and assumed $g_R=0.1$. Our new model ansatz generates large patterns of CP violation of ${\cal O}(1)$ above the open-charm threshold. Similar large effects were found in the recent analysis of \cite{Bediaga:2020ztp}, where a model based on hadronic charm loops combined with a $\chi_c^0$ resonance is discussed. Therefore, including these effects in the experimental analysis seems timely.

Of course, taking ${\cal A}_u$ constant is a large simplification, as in a full fit, this amplitude would contain a series of low-lying resonances (in fact also ${\cal A}_c$ would contain such resonances). To illustrate the effect that the interference of these resonances with ${\cal A}_c$ could have, we include a $\rho$ resonance parametrized by a Breit-Wigner shape in the amplitude ${\cal A}_u$. In Fig.~\ref{fig_dalitz_combi}, we show the corresponding CP distributions again for both the scalar $\chi_{c0}(3860)$ and the vector $X(3872)$ described by the resummed propagator $T_R$. Even though we parametrize the ${\cal A}_u$ in this simplistic way, the obtained distributions show a wide variety of CP violation in different regions of the Dalitz and the model generates ${\cal O}(1)$ CP violation in the high-energy region. To stress the difference of our model with a standard Breit-Wigner parametrization, we give in Fig.~\ref{fig_dalitz_BW} the corresponding plots using Breit-Wigner parametrizations for the $\rho$ and the $\chi_c$ and $X$ resonance. As the resonances are very narrow, CP violation is only observed in its vicinity. 


We conclude that our simple model ansatz to include effects from open-charm threshold states can indeed cause intricate structures with ${\cal O}(1)$ CP violation. Moreover, we suggest to perform the amplitude analysis in terms of the amplitudes ${\cal A}_u$ and ${\cal A}_c$, instead of ${\cal A}_+$ and ${\cal A}_-$. This allows to better distinguish effects from different physical sources and directly probe the strong phases, which need to be better understood. From the experimental point of view, once data becomes more precise, it will be relevant to include open-charm effects and exotic states like $\chi_{c0}(3860)$ or $X(3872)$, which can be easily done in our framework. 
Employing our new model ansatz into a full amplitude analysis may in fact give valuable insights and help improve the theoretical description of multibody decays not only in the high invariant mass region but also for the full phase space.

\section{Acknowledgements}
We would like to thank Tim Gershon, Ignacio Bediaga, Patricia Magalh\~aes, Jussara Miranda and Tobias Frederico for fruitful discussions. The  research  of  ThM and KO  has  been  supported  by the BMBF Forschungsschwerpunkt 
ErUM-FSP T04 under Grant No. 05H2018. The work of KKV is supported by the DFG Sonderforschungsbereich/Transregio 110 “Symmetries and theEmergence of Structure in QCD”.  We would like to thank to the Mainz Institute for Theoretical Physics (MITP) of the Cluster of Excellence PRISMA+ (Project ID 39083149) for its hospitality and support. KKV would like to especially thank Ignacio Bediaga, for hosting her at the Brazilian Center for Research in Physics (CBPF).

\bibliographystyle{JHEP} 
\bibliography{refs_Btopipipi.bib}
\end{document}